\documentclass[%
superscriptaddress,
notitlepage,
 amsmath,amssymb,
 aps,
prb,
]{revtex4-2}

\usepackage{graphicx, subfigure, caption}
\usepackage{dcolumn}
\usepackage{bm}
\usepackage{hyperref}
\usepackage[mathlines]{lineno}

\begin{document}


\title{Active resonator depletion with short microwave pulses}

\author{Sashank Kaushik Sridhar}
\affiliation{%
Indian Institute of Technology Madras, Chennai-600036, India
}%
\author{David P. DiVincenzo}%
 \email{d.divincenzo@fz-juelich.de}
\affiliation{%
 Peter Gr{\"u}nberg Institute (PGI-2), Forschungszentrum J{\"u}lich, D-52425, J{\"u}lich, Germany
}%



\date{\today}
\begin{abstract}
We propose a physical model to explain the phenomenon of photon depletion in superconducting microwave resonators in the dispersive regime, coupled to Josephson junction qubits, via short microwave pulses. We discuss the conditions for matching the amplitude and phase of the pulse optimally within the framework of the model, allowing for significant reductions in reset times after measurement of the qubits.
We consider how to deal with pulses and transient dynamics within the input-output formalism, along with a reassessment of the underlying assumptions for a wide-band pulse.
\end{abstract}

\maketitle


\section{\label{sec:intro}
Introduction}

Rapid measurements and rapid reset are crucial elements in the operation of superconducting quantum computers. Modern circuit-QED architectures largely rely on dispersive readout of qubits \cite{cqedintro}, i.e., measuring conditional shifts in the resonance peaks of microwave resonators dispersively coupled to the qubits. These superconducting microresonators typically have very high quality factors {\em Q} and low decay rates $\kappa$, due to the small value of the capacitive coupling to a Coplanar Waveguide (CPW) transmission line, also called a feedline \cite{strongcoup}. After measurement, the resonator must be depleted so as to not affect the qubit state in further operations (via the AC Stark shift \cite{cqedintro}). However, typical depletion times are limited by the same decay rate, leading to very slow passive reset. One proposal to tackle this is by sending microwave pulses in through the feedline to interact with the photons in the resonator and deplete it faster. The experiments \cite{Bultink2016,Gambetta2016} have been promising, although limited to an empirical understanding of the underlying physical processes. While there have been efforts to understand the numerical optimisation of the pulses better \cite{NumAnalysis2017}, our work focuses on establishing the theory and the equations governing the evolution of the systems.

We provide a two-pronged explanation for this phenomenon; a classical model and a quantum model, with perfect agreement between the two. Such an endeavour also acts as a foot-in-the-door to explore the often neglected ``short-pulse" regime. We also express what `photon depletion' in such a system really means, so as to root the theory on solid footing for further explorations. We find that the depletion time is entirely dependent on the pulse-width, and the amplitude and phase of the pulse can be fixed based on the derived matching conditions. Although limited to the dispersive approximation of the superconducting qubit-resonator system, the analytical expressions can be used as groundwork for further extensions that deal with nonlinear resonators and highly non-classical states of radiation.

\section{Proposed Setup}
\label{section:setup}

\begin{figure}
    \centering
    \includegraphics[width=\textwidth]{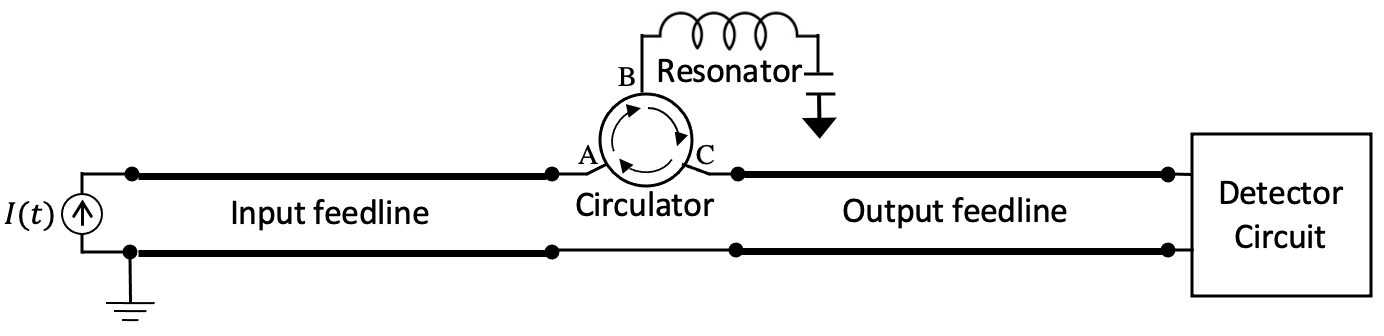}
    \caption{Schematic used in the model, with an arbitrary waveform generator acting as a time-varying source that is directly coupled to a finite transmission line. The output transmission line goes to a detector circuit and the required non-reciprocity is enforced by an ideal circulator coupled to the input, output and resonator nodes.}
    \label{fig:schematic}
\end{figure}

We assume a series LC-oscillator to model the linear response of a gap-coupled $\lambda/2$-transmission line resonator operating in its fundamental mode \cite{Pozar}, with the coupling strength controlled by adjusting the inductance or capacitance. The time-dependent voltage of this LC resonator would correspond to the peak voltage at the antinode in the transmission line resonator. The resonator is coupled to a superconducting qubit within the dispersive approximation with the Hamiltonian,
\begin{equation*}
    \hat{\mathcal{H}}_{disp}= \hbar\omega_r'\hat{a}^\dagger\hat{a}+\frac{\hbar\omega_q'}{2}\hat{\sigma}_z+\hbar\chi\hat{a}^\dagger\hat{a}\hat{\sigma}_z
\end{equation*}
where $\omega_r'$ and $\omega_q'$ are the effective resonator and qubit frequencies respectively, and $\chi$ is the dispersive shift (see Eq.~(44) in \cite{cqedreview}). When the qubit is measured, $\langle\hat{\sigma}_z\rangle$ collapses to $\pm 1$, thus the resonator frequency is  higher or lower than $\omega_r'$ by $\chi$ depending on the measured qubit state. Furthermore, the resonator coupled to the qubit is further capacitively coupled to the transmission lines via the circulator ports. We deal with these coupled entities in our model by treating their response as equivalent to a series LC-resonator with a resonant frequency that is determined by the qubit state, since the effect of the capacitive coupling to the feedline is also just a shift in the resonance peak \cite{Pozar}. This resonator is then connected to the B-port of a circulator \cite{Pozar}, with the input feedline on the A-port and an output feedline on the C-port (see Fig.~\ref{fig:schematic}). The input line is driven  with a time-varying current source, which is meant to represent the action of an arbitrary waveform generator. The output feedline carries the output signal from the resonator to a detector circuit that reads out the signal after amplification (usually done with a JPA \cite{JPA} followed by a high electron-mobility transistor).

\section{Classical Model}
\label{section:classical}

To understand the depletion phenomenon in the classical sense, we consider the voltage amplitude of the resonator node and posit that this must go to zero when depletion of photons occurs. This statement would become clearer when dealing with the quantum model.\\
In our analysis, we assume that the transmission line is excited in the TEM mode (practically quasi-TEM \cite{CPW}), which reduces the modelling to a 1D problem. We consider the flux $\Phi(x,t)$ as the main quantity of the transmission line, as the voltage and current at any given position and instant can be derived from partial derivatives of this flux variable. We assume $I(t)$ as the time-varying current from the source, $l(c)$ is the distributed inductance (capacitance) of the line and $v=1/\sqrt{lc}$ is the characteristic phase velocity of the line which has a linear dispersion relation in the TEM mode.

From the Telegrapher's equations of a lossless transmission line \cite{Pozar},
\begin{equation}
\label{eqn:tele}
    \frac{\partial^2\Phi(x,t)}{\partial t^2}=v^2\frac{\partial^2\Phi(x,t)}{\partial x^2}
\end{equation}
This gives a general solution consisting of a superposition of right and left-propagating parts,
\begin{equation*}
    \Phi(x,t)= \phi^+f^+(t-x/v)+\phi^-f^-(t+x/v)
\end{equation*}
Taking the current source end of the line as $x=0$ and the circulator port as $x=L_c$, we can assign boundary conditions to the flux.
\begin{equation}
\label{eqn:bc}
    \frac{\Phi'(0,t)}{l}= -I(t)
\end{equation}
where $\Phi'\equiv\frac{\partial\Phi}{\partial x}$ and $\dot{\Phi}\equiv\frac{\partial\Phi}{\partial t}$ are as per standard notations.\\
Assuming no residual excitations in the input feedline prior to turning on the current source, we then see that only the right-moving solution is allowed by the current source boundary condition due to causality.
\begin{equation}
\label{eqn:causal}
    I(t) = \frac{\phi^+}{vl}\frac{\mathrm{d}f^+}{\mathrm{d}t}(t-0/v)
    =\frac{\phi^+}{vl}\frac{\mathrm{d}f^+}{\mathrm{d}t}((t+x/v)-x/v)
    =-\frac{\Phi'(x,t+x/v)}{l}
    =I_{in}(x,t+x/v)
\end{equation}
Thus, we can uniquely define the waveform that propagates down the line from this boundary condition.\\
Here, we make the important assumption about the driving current that it has finite support in time,
\begin{equation}
\label{eqn:pulse}
    I(t)=0\,, \quad \text{if  } t<t_0-t_w \text{  or  } t>t_0
\end{equation}
where $t_0<0$ and $t_w$ is the pulse-width. We assume $t_0-t_w=-L_c/v$, so the pulse reaches the resonator at $t=0$.

The circulator acts as a directional element that blocks counter-propagation of signals, i.e., signals only go cyclically from port A to port B, B to C and C to A. The pulse is directed through the circulator to the resonator, where the boundary conditions depend on both the input and output signals,
\begin{gather}
    \label{eqn:Ib}
    I_B(t)=I_{in}(L_c,t)-I_{out}(L_c,t)\\
    \label{eqn:Vb}
    V_B(t)=V_{in}(L_c,t)+V_{out}(L_c,t)
\end{gather}
where $V_B(t)$ and $I_B(t)$ are the net voltage at port B and net current flowing into the resonator respectively.\\
The charge in the capacitor of the resonator can be obtained from the equation,
\begin{equation*}
    \ddot{q_r}=-\omega_r^2q_r-\frac{V_B(t)}{L_0}
\end{equation*}
where, $L_0$($C_0$) is the inductance (capacitance) of the resonator and $\omega_r=1/\sqrt{L_0C_0}$ is the resonant frequency.\\
Using the fact that $\dot{q_r}=I_B(t)$ and $V_{in}(x,t)=Z_0I_{in}(x,t)$ (similarly for $V_{out}(x,t)$), we get
\begin{gather}
    \label{eqn:qrdiff}
    \ddot{q_r}=-\omega_r^2q_r-\kappa\dot{q_r}-\frac{2V_{in}(L_c,t)}{L_0}\\
    \label{eqn:Vrdiff}
    \implies \ddot{V_r}=-\omega_r^2V_r-\kappa\dot{V_r}-2\omega_r^2Z_0I(t-L_c/v)
\end{gather}
where $\kappa=Z_0/L_0$ and $Z_0=\sqrt{\frac{l}{c}}$ is the characteristic impedance of the feedline.

Thus, the resonator behaves as a damped harmonic oscillator with a time-dependent drive. It is generally assumed that $\kappa<<\omega_r$ for high-Q resonators that are weakly out-coupled to the feedline.
\\This can now be solved by taking the Laplace transform:
\begin{equation*}
    s^2\Tilde{V}_r[s]+sV_r(0^-)-\dot{V_r}(0^-)= -\omega_r^2\Tilde{V}_r[s]+s\kappa\Tilde{V}_r[s]+\kappa V_r(0^-)-2\omega_r^2Z_0\Tilde{I}_{in}[s]
\end{equation*}
where $\Tilde{I}_{in}[s]$ is the Laplace transform of $I_{in}(L_c,t)$, and $I_{in}(L_c,t)=I(t-L_c/v)$. Note that the sign convention used here is opposite to what is the norm in most classical circuit-theory literature, so as to stay consistent with the physics convention of $e^{-i\omega t}$.

Simplifying this gives,
\begin{equation*}
    \Tilde{V}_r[s]=\frac{\frac{\kappa}{2} V_r(0^-)+\dot{V_r}(0^-)}{s^2-s\kappa+\omega_r^2}
    -\frac{(s-\frac{\kappa}{2})V_r(0^-)}{s^2-s\kappa+\omega_r^2}-2\omega_r^2Z_0\frac{\Tilde{I}_{in}[s]}{s^2-s\kappa+\omega_r^2}
\end{equation*}
Taking the limit $\kappa<<\omega_r$, the denominator reduces to $(s+\kappa/2)^2+\omega_r^2$, giving us the time-domain expression:
\begin{gather}
    \label{eqn:tdomain}
    V_r(t) = e^{-\kappa t/2}\Bigg[\left\{\frac{\kappa}{2\omega_r} V_r(0^-)+\frac{\dot{V_r}(0^-)}{\omega_r}\right\}\sin{\omega_r t}
    +V_r(0^-)\cos{\omega_r t}\Bigg]
    +i\omega_rZ_0\left(\int_0^t{\mathrm{d}\tau e^{(\frac{\kappa}{2}+i\omega_r)(\tau-t)}I_{in}(L_c,\tau)}\,-\,c.c\right)
    \\\implies V_r(t) = V_{nat}(t)+V_{dr}(t)
\end{gather}
\begin{figure}
    \centering
    \subfigure[]{\label{fig:amplitude}
    \includegraphics[width=.45\textwidth]{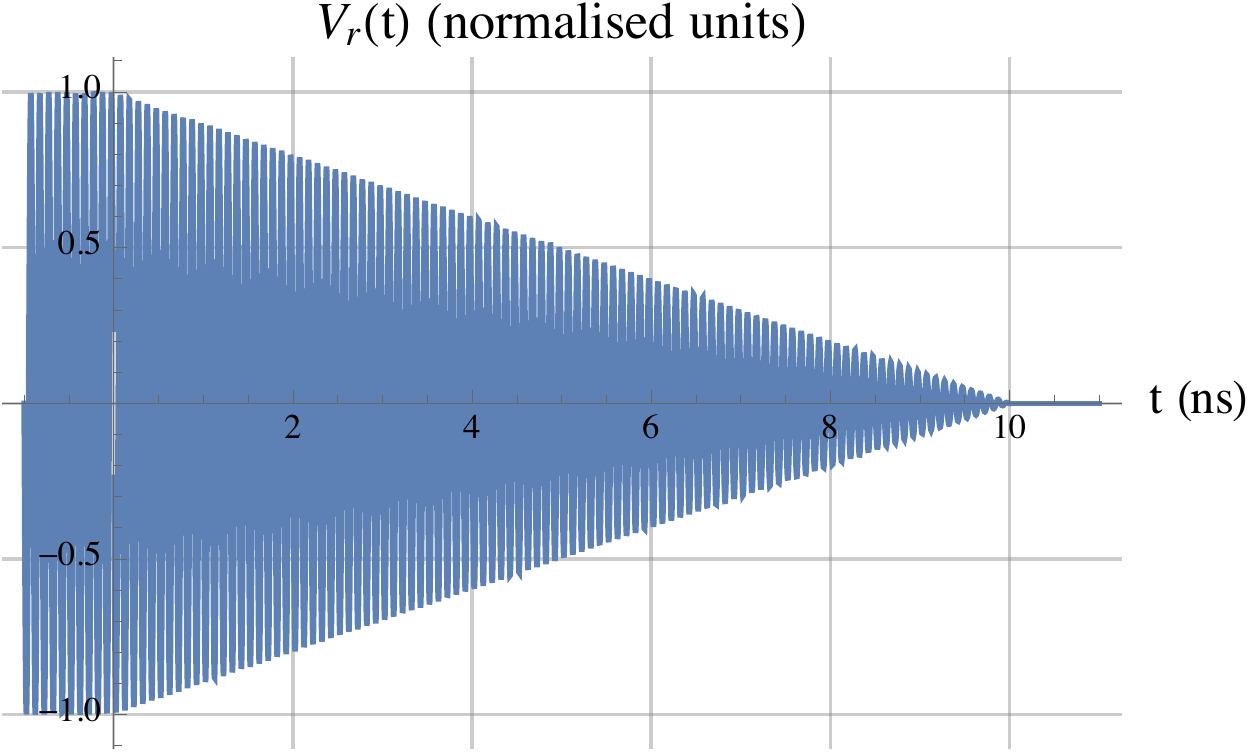}}
    \qquad
    \subfigure[]{\label{fig:phase}
    \includegraphics[width=.46\textwidth]{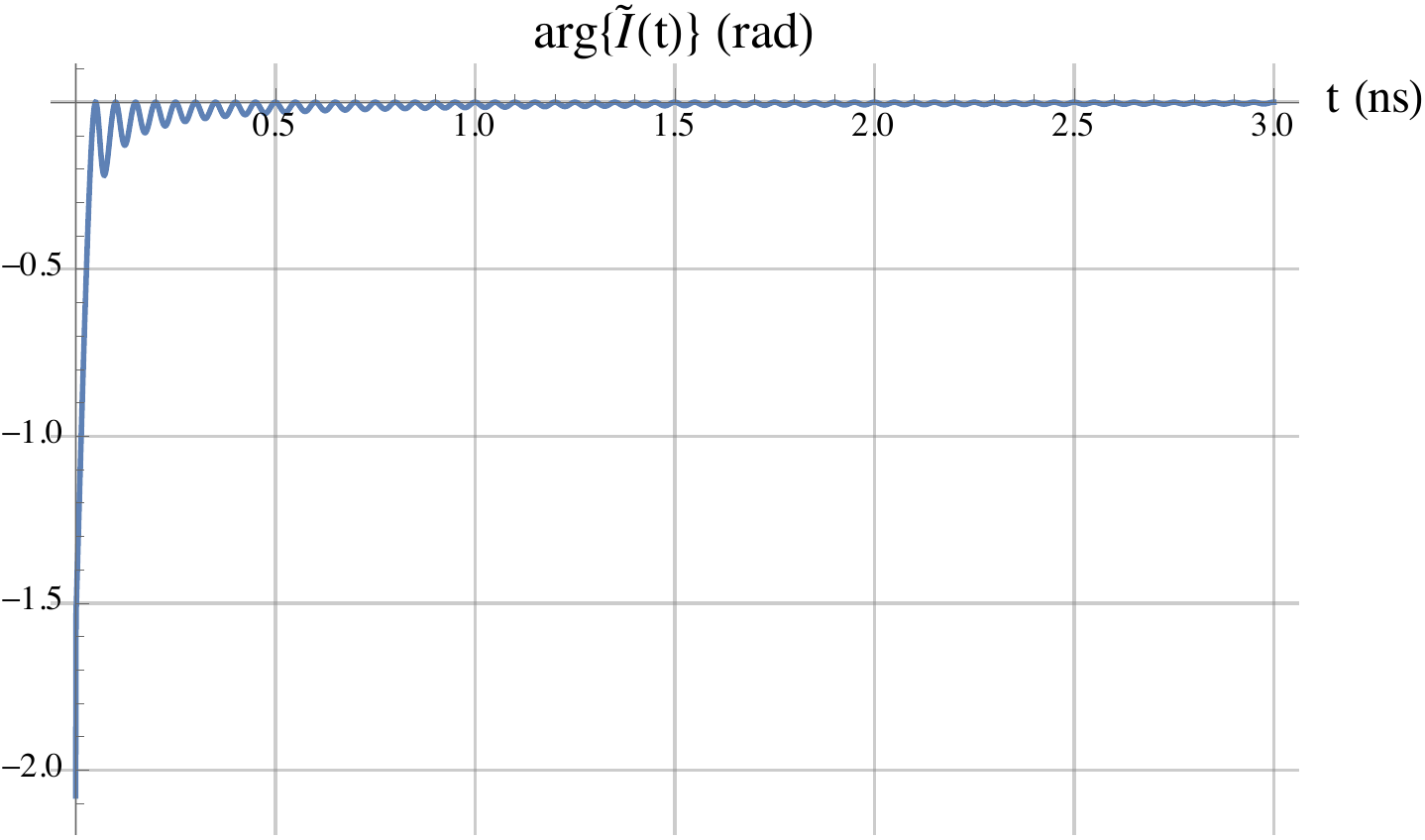}}
    \captionsetup{justification=centering}
    \caption{a) Transient voltage at the resonator node for the pulse $I_{in}(L_c,t)=0.5\sin(\omega_r t+\theta_p)$ for $0\,\mathrm{ns}<t<10\,\mathrm{ns}$, with $\omega_r=2\pi\times10\, \mathrm{GHz}$ and $\theta_p=\pi$, showing an almost linear drop till $t_w$ after which it stays at a constant amplitude, b) the time variation of $\arg\{\Tilde{I}[t]\}$ defined in Eq.~\eqref{eqn:I_tilde} for $\theta_p=\pi$, which settles to $0$ and is thus matched with a resonator voltage that has a phase of $(2n-1)\pi$ at $t=0\,\mathrm{ns}$.}
    \label{fig:transient}
\end{figure}
If we take the reference time $t_{fill}<0$ when the resonator is excited with photons from measuring the qubit(s), with an initial phase $\theta_{fill}$,
\begin{equation*}
    V_{nat}(t)= V_r(t_{fill})e^{-\kappa (t-t_{fill})/2}\sin{(\omega_r(t-t_{fill})+\theta_{fill})}
\end{equation*}
For $t\geq t_w$, we can write $V_{dr}(t)$ as:
\begin{equation*}
    V_{dr}(t) = -2\omega_r Z_0e^{-\kappa t/2}\mathrm{Im}\left(e^{-i\omega_rt}\Tilde{I}_{in}\left[s=\frac{\kappa}{2}+i\omega_r\right]\right)
    = 2\omega_r Z_0e^{-\kappa t/2}\left|\Tilde{I}_{in}\left[s=\frac{\kappa}{2}+i\omega_r\right]\right|\sin{(\omega_r t-\theta_I)}
\end{equation*}
where, $\theta_I=\arg\left\{\Tilde{I}_{in}\left[s=\frac{\kappa}{2}+i\omega_r\right]\right\}$.

By definition of the pulse in Eq.~\eqref{eqn:pulse}, $V_{dr}(t)$ attains a constant amplitude with an overall exponential decay at $t=t_w$. Beyond this point, if the amplitude of $V_{dr}(t)$ matches that of $V_{nat}(t)$ and the phases differ by $(2n-1)\pi$, then the voltage in the resonator goes to zero, i.e., the resonator gets depleted. This lets us define a set of 'matching' conditions:
\begin{gather}
    \label{eqn:matching}
    \left|\Tilde{I}_{in}\left[s=\frac{\kappa}{2}+i\omega_r\right]\right|=\frac{\exp{(\kappa t_{fill}/2)}}{2\omega_rZ_0}V_r(t_{fill})\\
    \theta_I=(2n-1)\pi+\omega_r t_{fill}-\theta_{fill}
\end{gather}
This gives a good physical picture of a high-Q resonator acting as a filter for the pulse, with an effective resonance peak at $\omega=\omega_r-i\kappa/2$. Furthermore, the matching conditions give a very clear picture of the protocol to follow. The resonator can initially be excited for a qubit read-out at $t_{fill}$, through the same input feedline and current source, with a weak microwave probe tone that is detected with a heterodyne measurement \cite{DSP}. It is possible to calculate and characterise the resultant amplitude and phase of the resonator voltage at the end of the protocol. One can thus send in the pulses that are matched to these quantities to fully deplete the photons. The transient dynamics for a particular choice of pulse are shown in Fig. \ref{fig:transient}.

\begin{figure}
    \centering
    \includegraphics[width=0.85\textwidth]{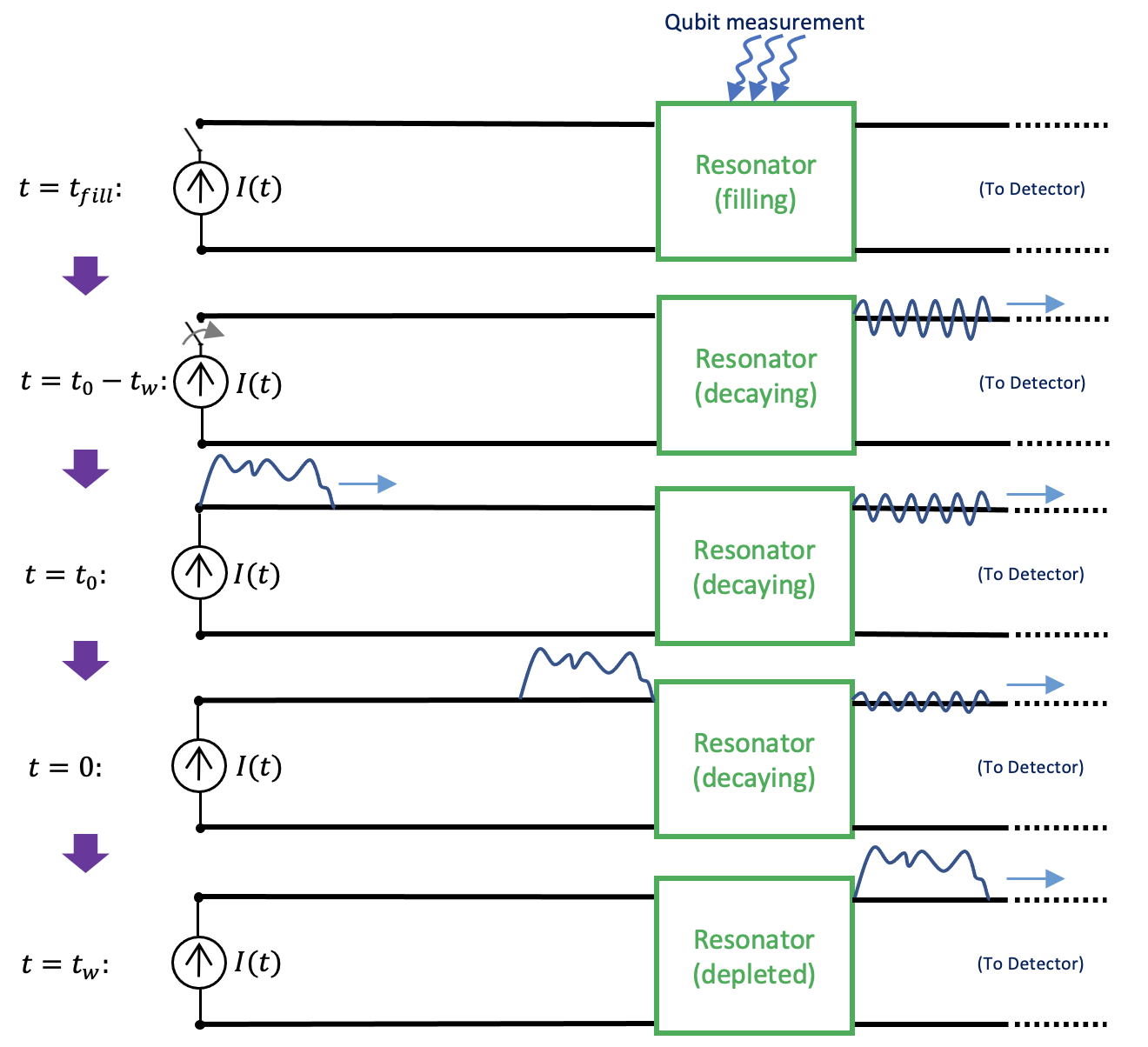}
    \caption{Schematic describing the relevant instants of time in our model. Note that the input and output lines are connected to the same port of the resonator via a circulator. At $t=t_{fill}$, the qubit(s) is(are) measured, exciting the resonator voltage. Subsequently, we see a decaying voltage from the resonator that is characterised by $\kappa$. At $t=t_0-t_w=-L_c/v$, the pulse is turned on and fully emitted by $t=t_0$. We set $t=0$ as the reference for when the pulse is incident on the resonator, and the interaction occurs till $t_w$, the pulse width. The output waveform is as given in Section~\ref{subsection:output}.}
    \label{fig:pulseprop}
\end{figure}

\subsection{Output Signal}
\label{subsection:output}
One can verify the occurence of the depletion by detecting the output signal, calculated from Eq.~\eqref{eqn:Ib},
\begin{equation}
    I_B(t)=I_{in}(L_c,t)-I_{out}(L_c,t)=C_0\dot{V_r}(t)
\end{equation}
From Eq.~\eqref{eqn:tdomain},
\begin{equation*}
    \dot{V_r}(t)=-\frac{\kappa}{2}V_r(t)
    +\omega_r^2\Bigg[\frac{V_r(t_{fill})}{\omega_r}e^{\kappa (\Delta t)/2}\cos{(\omega_r(\Delta t)+\theta_{fill})}
    +Z_0\left(\int_0^t{\mathrm{d}\tau e^{(\frac{\kappa}{2}+i\omega_r)(\tau-t)}I(\tau-L_c/v)}+c.c\right)\Bigg]
\end{equation*}
where $\Delta t=t-t_{fill}$.
We define, 
\begin{equation}
\label{eqn:I_tilde}
    \Tilde{I}[t]=\int_0^t{\mathrm{d}\tau e^{(\frac{\kappa}{2}+i\omega_r)\tau}I(\tau-L_c/v)}
\end{equation}
\begin{eqnarray*}
    \therefore \dot{V_r}(t)&=&-\frac{\kappa}{2}V_r(t)+\omega_r\Bigg[V_r(t_{fill})e^{\kappa (\Delta t)/2}\cos{(\omega_r(\Delta t)+\theta_{fill})}
    +2Z_0\omega_r\big|\Tilde{I}[t]\big|e^{-\kappa t/2}\cos{\left(\omega_rt+\arg\{\Tilde{I}[t]\}\right)}\Big]\\
    &=&\omega_r\Bigg[V_r(t_{fill})e^{\kappa (\Delta t)/2}\cos{(\omega_r(\Delta t)+\theta_{fill}+\theta_r)}
    +2Z_0\omega_r\big|\Tilde{I}[t]\big|e^{-\kappa t/2}\cos{\left(\omega_rt+\arg\{\Tilde{I}[t]\}+\theta_r\right)}\Bigg]
\end{eqnarray*}
where $\theta_r=\sin^{-1}\left(\frac{\kappa}{2\omega_r}\right)$. Therefore, the out-flowing signal is given by:
\begin{gather*}
    V_{out}(L_c,t)\,(0\leq t\leq t_w)=Z_0\left(I(t-L_c/v)-C_0\dot{V_r}(t)\right)\\= Z_0I(t-L_c/v)
    -\frac{\kappa}{\omega_r}\left[V_r(t_{fill})e^{\kappa (\Delta t)/2}\cos(\omega_r(\Delta t)+\theta_{fill}+\theta_r)+2Z_0\omega_r \big|\Tilde{I}[t]\big|e^{-\kappa t/2}\cos\left(\omega_rt+\arg\{\Tilde{I}[t]\}+\theta_r\right)\right]
\end{gather*}
The output would show a weak, underdamped sinusoidal oscillation from the photons present in the resonator prior to the pulse reaching it, followed by the reflected pulse and no signal after that (amplitudes cancel out from the matching conditions). The output contains a phase-shifted version of the voltage transient of the resonator, which gives a clear indication of whether the depletion has occured or not. The coefficient of $\kappa/\omega_r$ however leads to a very weak contribution from the resonator output as compared to the reflected pulse, so the output signal must be analysed by subtracting out the input pulse.

It is important to note that the same results can be derived by solving for the normal modes of the transmission line. We use this very technique in the next section to derive the photon picture, so we avoid mentioning it in the classical model.

\subsection{Motivation for Quantum Model}
\label{subsection:motivation}
From the classical theory, depletion can be explained under the high-Q cavity limit as interference between the pulse component that gets filtered into the resonator and the standing wave mode present within it. This model can already give us rich insights about the phenomenon and related observables, but it is still unclear how the voltage amplitude connects to the photon number depletion. Moreover, the model would be inherently insufficient in explaining the depletion of photons in squeezed states of the resonator, which would be useful for precision readout \cite{Dispersive2014}. Thus, a more general quantum model would serve to generalise and broaden the scope, further allowing one to potentially probe the classicality of the radiation states in the resonator.

\section{Quantum Model}
\subsection{Quantisation in CQED}
\label{section:Quantisation}

In vacuum quantum electrodynamics \cite{QED}, the EM fields are spatially represented in a Fourier basis, with the basis functions reinterpreted as field operators and represented in terms of creation and annihilation operators. The multimode EM radiation can then be viewed as an 
ensemble of photons \cite{Titulaer66}. These spatial basis modes change depending on the system properties and geometric constraints, such as in cavity QED \cite{cQED_Becker, Kimble_1998}.

We utilise the method for quantisation as stated in \cite{Wendin} and \cite{Vool_2017}, and use \cite{Gardiner} and the Appendix of \cite{QNoise} as a reference for the Input-Output Theory. Recent work on Input-Output theory with quantum pulses \cite{Molmer} provides motivation for a generalised framework to accommodate multi-mode quantum radiation and its interaction with classical and quantum scatterers in the presence of different loss channels. Along similar lines, we show the treatment of short pulses in a superconducting transmission line as multi-mode coherent states. A detailed discussion of the photon number in these short pulses can be found in Appendix~\ref{section:Appendix}.

\begin{figure*}[t]
    \centering
    \includegraphics[width=0.8\textwidth]{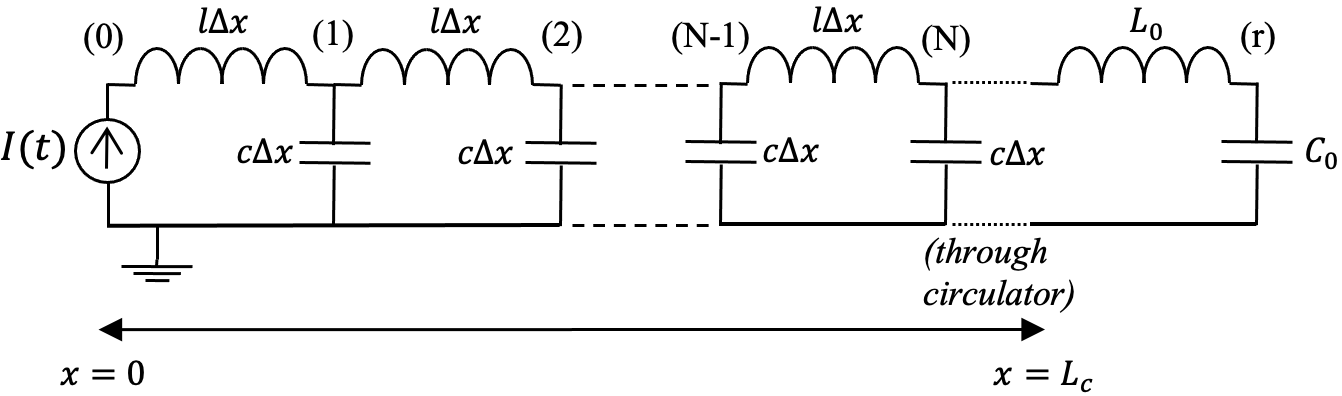}
    \captionsetup{justification=centering}
    \caption{Discretised representation of a lossless transmission line with infinite coupled LC oscillators. The inductances and capacitances are taken as differential elements that depend linearly on their size $\Delta x$, which goes to $0$ in the continuum limit, while $N\to \infty$, and it is ultimately coupled to the readout resonator, which is modeled by a finite inductance $L_0$ and capacitance $C_0$. The flux variable for each node is marked by the node index in parentheses.}
    \label{fig:disc_line}
\end{figure*}
First we will consider the system only till the $N^{th}$ node of the line, without considering any coupling to the resonator. With the Kirchhoff current equations at each node acting as the Hamilton equations of motion, we can construct the system Hamiltonian,
\begin{equation}
    \label{eqn:H_trans}
    \mathcal{H}=\sum_{m=1}^N{\frac{q_m^2}{2c\Delta x}}+\sum_{m=2}^N{\frac{(\phi_m-\phi_{m-1})^2}{2l\Delta x}}-I(t)\phi_1
\end{equation}
where $(\phi_i,q_i)$ are canonically conjugate variables of the flux and charge at each node. The summations can be rewritten as matrices, with the 
inverse capacitance matrix proportional to $\mathbb{I}$. The inverse inductance matrix is symmetric and is guaranteed to be diagonalisable.\\
After diagonalisation,
\begin{eqnarray}
    \label{eqn:H_diag}
    \mathcal{H}&=&\mathcal{H}_0+\sum_{j=1}^{N-1}{\left(\frac{\tilde{q}_j^2}{2c\Delta x}+\frac{c_j\tilde{\phi}_j^2}{2l\Delta x}-I(t)\mathbf{O}_{j1}\tilde{\phi}_j\right)}\\
    &=& \mathcal{H}_0+\sum_{j=1}^{N-1}{\mathcal{H}_j}\\
    \label{eqn:H_dc}
    \mathcal{H}_0&=&\frac{\tilde{q}_0^2}{2c\Delta x}-I(t)\frac{\tilde{\phi}_0}{\sqrt{N}}\\
    \omega_j&=&\frac{\sqrt{c_j}}{\Delta x\sqrt{lc}}=\frac{v\sqrt{c_j}}{\Delta x}=\frac{v}{L_c}(N\sqrt{c_j})=vk_j
\end{eqnarray}
where, $c_j$ is the eigenvalue of the inverse inductance matrix, $(\tilde{\phi}_j,\tilde{q}_j)$ are the transformed canonically conjugate variables of the normal modes and $\mathbf{O}_{jm}$ is the element of the orthogonal matrix which implements the canonical transformation $\overrightarrow{\mathbf{\tilde{\phi}}}=\mathbf{O}\,\overrightarrow{\phi}$, where $\overrightarrow{\phi}$ and $\overrightarrow{\mathbf{\tilde{\phi}}}$ are the vectors of flux variables at each node and in each normal mode, respectively.

\begin{figure}
    \centering
    \subfigure[]{
    \label{fig:eigval}
    \includegraphics[width=.45\textwidth]{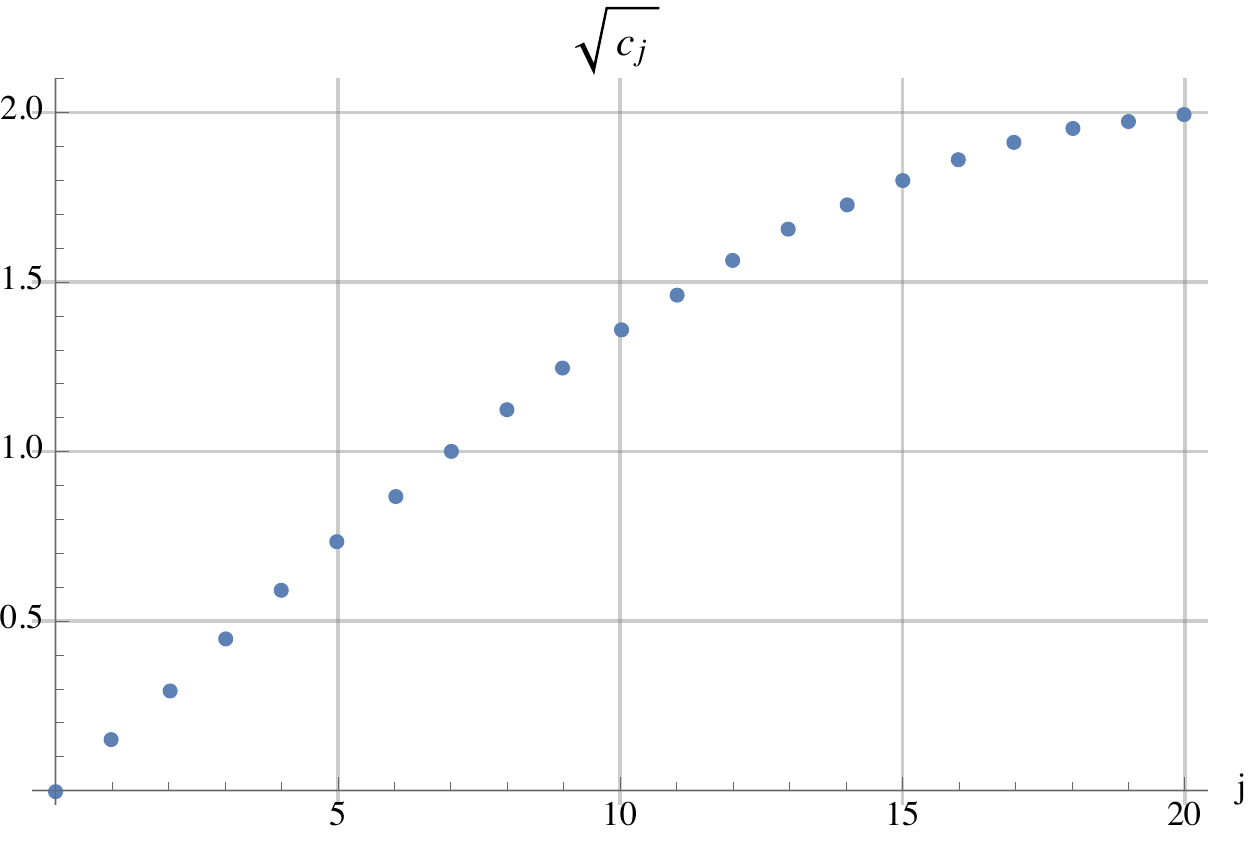}}
    \subfigure[]{
    \label{fig:eigmode1}
    \includegraphics[width=.45\textwidth]{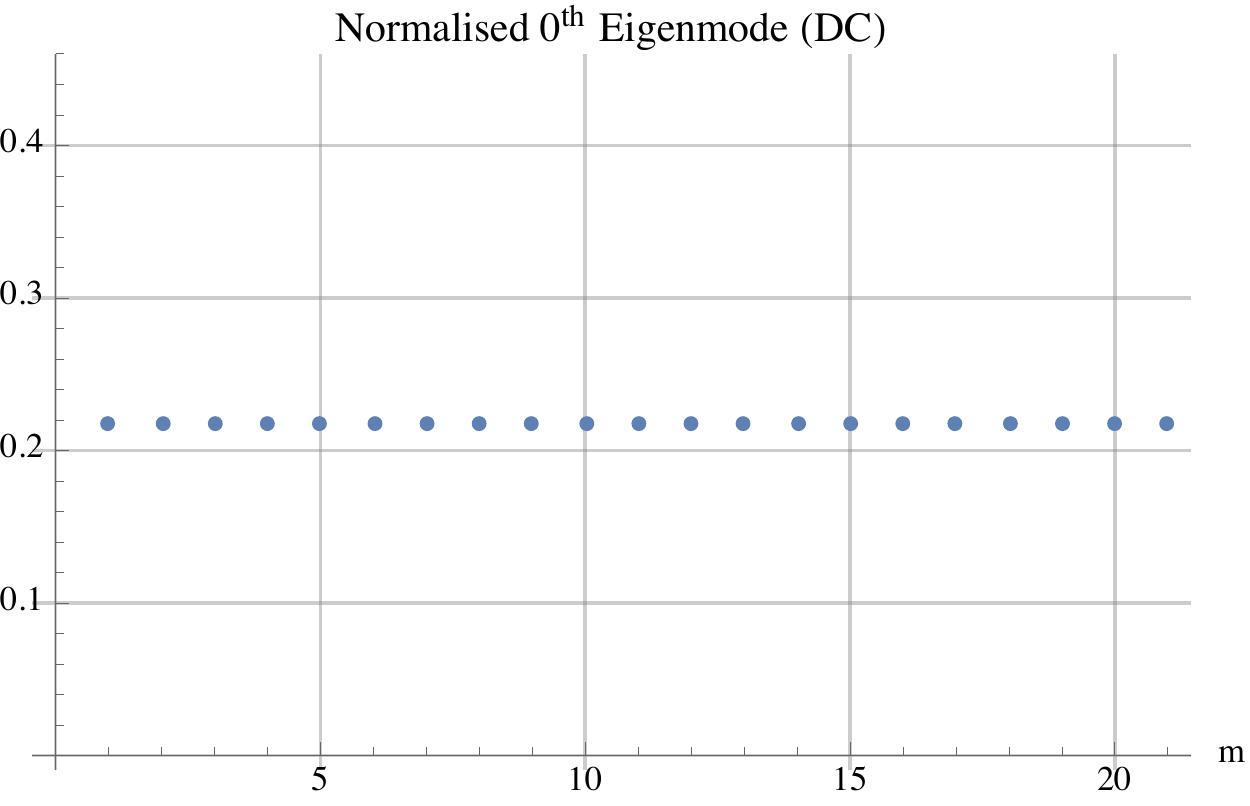}}
    \subfigure[]{
    \label{fig:eigmode2}
    \includegraphics[width=.45\textwidth]{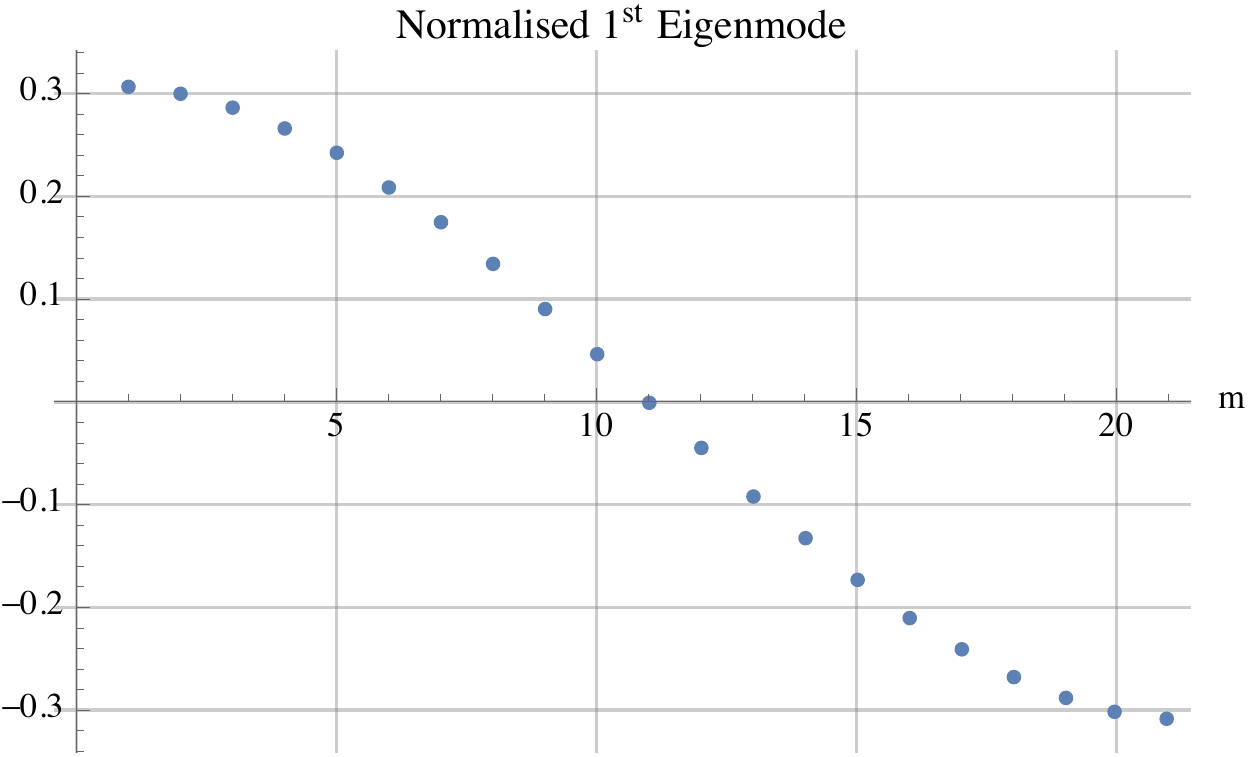}}
    \subfigure[]{
    \label{fig:eigmode3}
    \includegraphics[width=.45\textwidth]{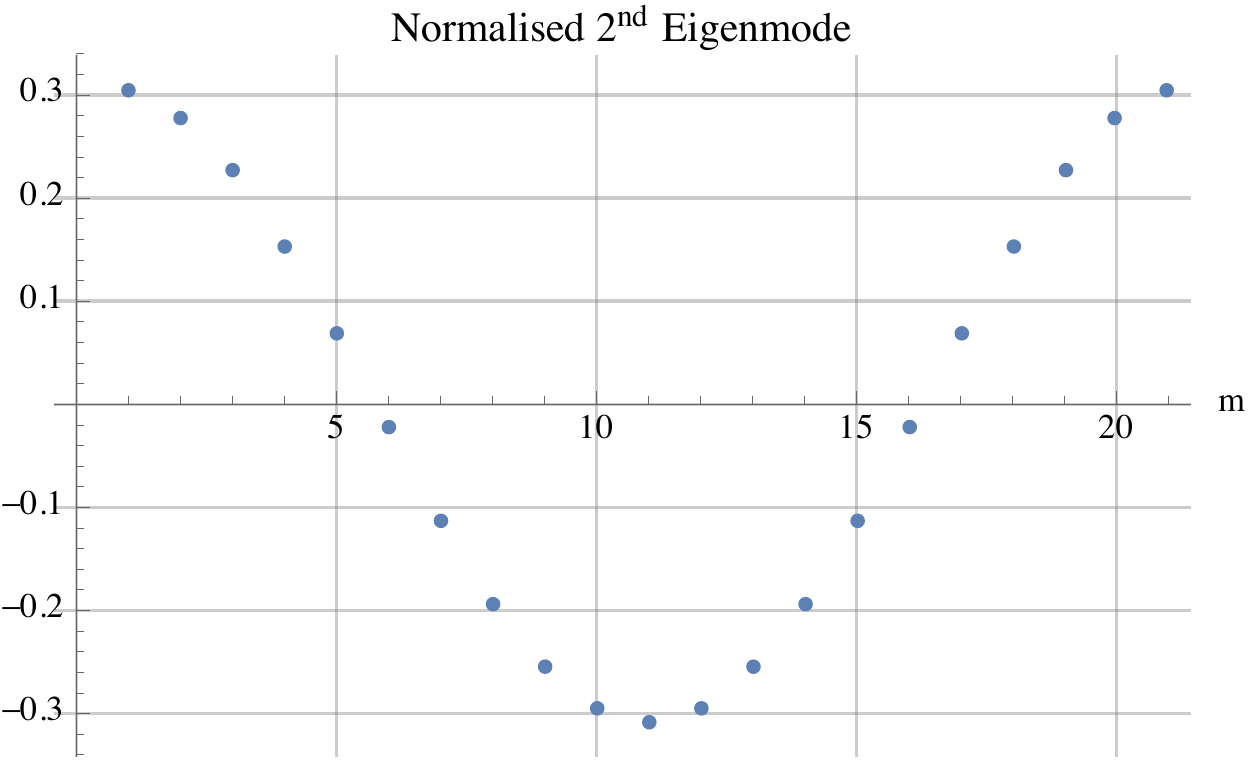}}
    \caption{a) Square root of the eigenvalue of each normal mode which is proportional to the normal mode frequency and shows characteristic behaviour of periodic systems, b) The normalised eigenmode with an eigenvalue $0$, corresponding to the ``DC'' mode which is a normalised sum over all the node variables, whose EOM provides a statement of charge/flux conservation; c) and d) are the first and second eigenmodes of the transmission line, where one can see the standing waves even in the discretised line. All data is plotted for $N=21$.}
    \label{fig:normal}
\end{figure}

Taking the continuum limit for our discrete model, i.e., $N\to\infty$ and $\Delta x\to 0$ such that $m\Delta x=x$ and $N\Delta x=L_c$,
\begin{eqnarray*}
    \mathbf{O}_{jm}&=&O_j\cos(jm\pi/N)\approx\sqrt{\frac{2}{N}}\cos\left(\frac{jx\pi}{L_c}\right)\\
    \sqrt{c_j}&\approx&\frac{j\pi}{N}\implies k_j=\frac{j\pi}{L_c}\\
    \therefore \mathbf{O}_{jm}&\approx& \sqrt{\frac{2}{N}}\cos(k_jx)
\end{eqnarray*}
Numerics for $N=21$ in Fig.~\ref{fig:normal} gives results in line with the continuum limit. We show
the square-root of the eigenvalue spectrum (proportional to the mode frequencies) and the first three eigenmodes, ordered by the node indices. The zeroth mode corresponds to the DC term, i.e., the sum over all the charges or fluxes of the line. Its contribution 
in the following can be neglected by choosing the pulse such that its time integral goes to zero.
Fig. \ref{fig:eigval} shows an initial linear behaviour, beyond which it tapers out. This behaviour is typical of periodic systems with discrete translational symmetries, like cubic lattices. When taking a continuum limit, the linear regime becomes the relevant part of the spectrum as $\omega_{N-1}\to \infty$.

We can now diagonalize the Hamiltonian by defining the creation and annihilation operators in the normal mode basis:
\begin{gather}
    \tilde{\phi}_j (j\geq 1)=\sqrt{\frac{\hbar Z_0}{2\sqrt{c_j}}}\,(\hat{a}_j+\hat{a}_j^\dagger)\\
    \tilde{q}_j (j\geq 1)=\frac{1}{i}\sqrt{\frac{\hbar \sqrt{c_j}}{2Z_0}}\,(\hat{a}_j-\hat{a}_j^\dagger)\\
    \therefore \mathcal{H}_j=\hbar\omega_j(\hat{a}_j^\dagger\hat{a}_j+1/2)-\sqrt{\frac{\hbar Z_0\omega_c}{\pi\omega_j}}I(t)(\hat{a}_j+\hat{a}_j^\dagger)
\end{gather}
This represents a multi-mode harmonic oscillator with a linear drive term, perfectly analogous to a multi-mode driven cavity \cite{Kimble_1998}. The driving can be represented by a displacement operator, giving rise to a coherent state in each mode of the transmission line.
\begin{gather}
    \label{eqn:drive_trans}
    \dot{\hat{a}}_j(t)=i\omega_j\hat{a}_j+i\sqrt{\frac{Z_0\omega_c}{\pi\hbar\omega_j}}I(t)\hat{\mathbb{I}}\\
    \implies \langle\hat{a}_j(t)\rangle=\alpha_j(t)=\sqrt{N}C_j(t)=i\sqrt{\frac{Z_0\omega_c}{\pi\hbar\omega_j}}e^{-i\omega_jt}\int_{t_0-t_w}^t{\mathrm{d}\tau e^{i\omega_j\tau}I(\tau)}\\
    \implies \alpha_j(t)=\alpha_j(t_0)e^{-i\omega_j(t-t_0)}\quad(t\geq t_0)
\end{gather}
where $t_0$ and $t_w$ have been defined in \eqref{eqn:pulse}, $C_j$ is the normalised coefficient of the annihilation operator of each mode (refer to Eq.~\eqref{eqn:mode_coeff}) and $N$ is the total average photon number of the multi-mode coherent state.

Thus, at $t=t_0$ the pulse is fully formed, and the multimode coherent state amplitude is fixed and evolves freely under the Harmonic oscillator Hamiltonian until $t=0$, after which the interaction with the resonator takes place.

\subsection{Coupling to resonator}
With the pulse displacing the transmission line modes to a multi-mode coherent state, we can now write the full Hamiltonian with the coupling to the resonator:
\begin{gather}
    \label{eqn:H_transreso}
    \mathcal{H}=\mathcal{H}_0+\sum_{j=1}^{\infty}\left(\mathcal{H}_j+\mathcal{H}_{coup,j}\right)+\mathcal{H}_r\\
    \label{eqn:H_coup}
    \sum_{j=1}^{\infty}\mathcal{H}_{coup,j}=-\sqrt{\frac{2}{N}}\frac{\phi_r}{L_0}\sum_{j=1}^{\infty}(-1)^j\tilde{\phi}_j=-\frac{\phi_r\Phi(L_c,t)}{L_0}\\
    \label{eqn:H_r}
    \mathcal{H}_r=\frac{\phi_r^2}{2L_0}+\frac{q_r^2}{2C_0},\, \omega_r=\frac{1}{\sqrt{L_0C_0}}
\end{gather}
Thus, we can see that there is a coupling term between the resonator mode and every mode of the transmission line, with the $(-1)^j$ term coming from the parity of the eigenmode at the end of the line ($\mathbf{O}_{jN}=(-1)^j\sqrt{2/N}$ ). The complete Hamiltonian would also include the coupling to the output feedline modes, but we neglect that in our expression for brevity, keeping the coupling bidirectional here and enforcing the non-reciprocity in the equations of motion (an in-depth treatment of non-reciprocal elements can be found in \cite{ParraRodriguezDiVincenzo}).
\\\\
Writing these in second quantization:
\begin{gather}
    \label{eqn:H_r_q}
    \mathcal{H}_r=\hbar\omega_r(\hat{a}_r^\dagger\hat{a}_r+1/2)\\
    \label{eqn:H_coup_q}
    \mathcal{H}_{coup,j}=-\hbar(-1)^j\sqrt{\frac{Z_0\omega_r\omega_c}{2\pi\omega_jL_0}}(\hat{a}_j+\hat{a}_j^\dagger)(\hat{a}_r+\hat{a}_r^\dagger)
\end{gather}
We can simplify the Hamiltonian by taking the Rotating Wave Approximation for $\mathcal{H}_{coup,j}$. This is justified by Eq.~\eqref{eqn:kappa_classical}; the resonator loss rate $\kappa$ goes inversely as the inductance $L_0$ of the LC resonator in our model, which can be tuned to satisfy $\kappa<<\omega_r$ in the dispersive approximation.
\begin{eqnarray}
    \label{eqn:H_coup_rwa}
    \mathcal{H}_{coup,j}&=&-i\hbar(f_j\hat{a}_j\hat{a}_r^\dagger-f_j^*\hat{a}_j^\dagger\hat{a}_r)\\
    \label{eqn:f_j}
    f_j&=&-i(-1)^j\sqrt{\frac{Z_0\omega_r\omega_c}{2\pi\omega_jL_0}}
\end{eqnarray}
The key feature is that the coupling terms come into the picture only when $\langle\Phi(L_c,t)\rangle$ is non-zero. 
Despite the photon states of each mode being excited at $t=t_0-t_w$, they do not interact with the resonator modes until $t=0$ due to causality (see Eq.~(E58) in the appendices of \cite{QNoise}). However, the resonator mode starts to interact with the modes of the output feedline as soon as the measurement happens at $t_{fill}$. A visual representation of this can be viewed in Fig.~\ref{fig:pulseprop}.

\subsection{Depletion of photons from the resonator mode}
We define $\hat{b}_j$ to be the annihilation operator of the standing wave modes in the output feedline, distinct from $\hat{a}_j$. The coupled equations of motion for the second quantised operators are given by, 
\begin{gather}
    \label{eqn:reso_EOM}
    \dot{\hat{a}}_r=-i\omega_r\hat{a}_r- \sum_{j=1}^{\infty}f_j\hat{b}_j\quad(t_{fill}\leq t<0)\\
    \label{eqn:out_EOM}
    \hat{b}_j(t)=f_j^*e^{-i\omega_jt}\int_{t_{fill}}^t\mathrm{d}\tau e^{i\omega_j\tau}\hat{a}_r(\tau)
\end{gather}
\begin{equation}
\label{eqn:decay_EOM}
    \therefore \dot{\hat{a}}_r=-i\omega_r\hat{a}_r
    -\sum_{j=1}^{\infty}|f_j|^2\int_{t_{fill}}^t\mathrm{d}\tau e^{-i(\omega_j-\omega_r)(t-\tau)}[e^{i\omega_r(t-\tau)}\hat{a}_r(\tau)]
\end{equation}
We can now define the loss rate of the cavity $\kappa$ from Fermi's Golden rule, as per Eq.~E24 in \cite{QNoise}.
\begin{equation}
\label{eqn:FGR}
    \kappa(\omega_r) = 2\pi\sum_j|f_j|^2\delta(\omega_r-\omega_j)
\end{equation}
The frequency dependence of $\kappa$ can be neglected under the Markov approximation \cite{walls2012quantum}, supported further by the restriction to an LC-resonator with a single resonant mode. From this, we end up with
\begin{equation}
\label{eqn:Markov}
    \sum_{j=1}^{\infty}|f_j|^2e^{-i(\omega_j-\omega_r)(t-\tau)}=\kappa\delta(t-\tau)
\end{equation}
Plugging this into Eq.~\eqref{eqn:decay_EOM},
\begin{equation}
    \dot{\hat{a}}_r=-(i\omega_r+\kappa/2)\,\hat{a}_r(t)
\end{equation}
When the pulse is incident on the resonator,
\begin{equation}
\label{eqn:reso_EOM_drive}
    \dot{\hat{a}}_r = -(i\omega_r+\kappa/2)\,\hat{a}_r(t)-\sum_{j=1}^{\infty}f_j\hat{a}_j(t)\quad(0\leq t<t_w)
\end{equation}
\begin{equation}
\label{eqn:reso_long}
    \begin{split}
        \implies \langle \hat{a}_r(t)\rangle=\alpha_r(t)=\,&e^{-(i\omega_r+\kappa/2)t}\alpha_r(0^-)\\
        -\,&e^{-(i\omega_r+\kappa/2)t}\sum_{j=1}^{\infty}f_je^{-i\omega_jL_c/v}\int_0^t\mathrm{d}\tau e^{(i\omega_r+\kappa/2)\tau}\left(i\sqrt{\frac{Z_0\omega_c}{\pi\hbar\omega_j}}\right)e^{-i\omega_j\tau}\left(\int_0^{t_w}\mathrm{d}\tau'e^{i\omega_j\tau'}I(\tau'-L_c/v)\right)
    \end{split}
\end{equation}
We define $s_0=i\omega_r+\kappa/2$. Using (31) and the fact that $e^{-i\omega_jL_c/v}=e^{-ij\pi}=(-1)^j$,
\begin{equation}
\label{eqn:reso_short}
    \alpha_r(t)=e^{-s_0t}\left[\alpha_r(0^-) -\sqrt{\frac{2L_0}{\hbar\omega_r}}\int_0^t\mathrm{d}\tau e^{s_0\tau} \int_0^{t_w}\mathrm{d}\tau'I(\tau'-L_c/v)\sum_{j=1}^{\infty}|f_j|^2e^{-i\omega_j(\tau-\tau')}\right]
\end{equation}
Substituting Eq.~\eqref{eqn:Markov},
\begin{equation}
\label{eqn:reso_final}
    \alpha_r(t)=e^{-s_0t}\left[\alpha_r(0^-) -\kappa\sqrt{\frac{2L_0}{\hbar\omega_r}}\int_0^{t}\mathrm{d}\tau e^{s_0\tau}I(\tau-L_c/v)\right]=e^{-s_0t}\left[\alpha_r(0^-) -\kappa\sqrt{\frac{2L_0}{\hbar\omega_r}}\tilde{I}(t)\right]
\end{equation}
Thus at $t=t_w$, for full depletion:
\begin{equation}
\label{eqn:photon_match}
    \alpha_r(0^-)=\kappa\sqrt{\frac{2L_0}{\hbar\omega_r}}\tilde{I}_{in}\left[s=\frac{\kappa}{2}+i\omega_r\right]
\end{equation}
Depletion is thus shown to occur within the generalised Input-Output framework, as is elucidated by the matching condition defined above.

\subsection{Consistency with Classical picture}
We now possess the required information to justify our original claim that photon depletion in the classical model can be viewed as the drop in the voltage amplitude of the resonator node.\\
From the quantisation of the resonator mode, we can get $\langle q_r(t)\rangle$, and from $\langle q_r(t)\rangle=C_0\langle V_r(t)\rangle$,
\begin{equation}
\label{eqn:Vr_expectation}
    \langle V_r(t)\rangle = \sqrt{\frac{2\hbar\omega_r}{C_0}}\mathrm{Im}\{\alpha_r(t)\}
\end{equation}
Rewriting $|\alpha_r(0^-)|$ as $|\alpha_r(t_{fill})|e^{\kappa t_{fill}/2}$ and defining $\theta_{fill}=\pi-\arg\{\alpha_r(t_{fill})\}$,
\begin{equation*}
    \langle V_r(t)\rangle= \sqrt{\frac{2\hbar\omega_r}{C_0}}|\alpha_r(t_{fill})|e^{-\kappa(t-t_{fill})/2}\sin(\omega_r(t-t_{fill})+\theta_{fill})
    +2\kappa\sqrt{\frac{L_0}{C_0}}|\tilde{I}(t)|\sin(\omega_rt-\arg\{\tilde{I}(t)\})
\end{equation*}
\\Using Eq.~\eqref{eqn:FGR} and taking the approximation $\omega_c<<\omega_r$,
\begin{equation}
\label{eqn:kappa_classical}
    \begin{split}
        \kappa &= \sum_{j=1}^\infty\frac{\omega_c\omega_rZ_0}{\omega_jL_0}\delta(\omega_r-\omega_j)\\
        &\approx \frac{\omega_rZ_0}{L_0}\int_0^\infty\mathrm{d}\omega\frac{\delta(\omega_r-\omega)}{\omega}\\
        &=\frac{Z_0}{L_0}
    \end{split}
\end{equation}
Substituting Eq.~\eqref{eqn:kappa_classical},
\begin{equation}
\label{eqn:Vr_final_exp}
    \langle V_r(t)\rangle = V_r(t_{fill})e^{-\kappa(t-t_{fill})/2}\sin(\omega_r(t-t_{fill})+\theta_{fill})+2Z_0\omega_re^{-\kappa t/2}|\tilde{I}(t)|\sin(\omega_rt-\arg\{\tilde{I}(t)\})
\end{equation}
Thus, we recover the classical equation for the resonator voltage, along with perfect agreement in the loss rate of the resonator. This establishes the self-consistency of the model and allows further extensions to non-classical regimes.

\subsection{The short-pulse regime}

A salient feature of the wide-band Input-Output theory is the coupling coefficient $f_j$ not being taken as a constant, as shown in Eq.~\eqref{eqn:f_j}. 
It is shown to vary as $\sqrt{\frac{1}{\omega_j}}$ along with a $(-1)^j$ factor arising from the parity at the $N^{th}$ node for each eigenmode. 
It must be noted that the nature of this coupling coefficient would change if instead of the series LC-resonator in Fig.~\ref{fig:schematic} and Fig.~\ref{fig:disc_line}, we model the resonator as a parallel LC-resonator that is capacitively coupled to the transmission line, leading to a proportionality to $\sqrt{\omega_j}$ which causes a divergence in the total photon number of the pulse. This divergence has been studied in depth \cite{Parra_Rodriguez_2018} and is dealt with by looking at the exact nature of the canonical transformation and thereby defining an ultraviolet cutoff for the modes under consideration. Moreover, the divergence in zero point fluctuations for this multi-mode system is resolved when accounting for the detectors having finite bandwidths. 

As seen in the classical model and its correspondence with the quantum picture, the relevant observables are the voltage operators of the transmission lines and the resonator. One can analyse the power spectral density of the voltage pulse and follow the steps in Appendix~\ref{section:Appendix} to calculate the average number of photons per mode. The question then arises - where are these photons? 

When a microwave pulse is mixed with an LO and low-pass filtered, the pulse's central frequency gets downshifted to allow the ADC to convert and detect the signal. The bandwidth of the detector plays a key role in characterising the quantum noise and accurately detecting the signal. Usually, this is given by the sampling rate of the ADC, which needs to be at least twice the bandwidth of the pulse to detect it accurately (from the Shannon-Nyquist Sampling Theorem). The idea of time-resolved photon counting begins to break down for such broadband detectors, as the expression for the photon flux $\phi=P/\hbar\omega_{central}$ no longer remains valid. 

We can, however, conclude that the off-resonant photons lie strictly within the pulse because of causality. This is easily demonstrated by calculating a Fourier Transform of the detected signal at different time-steps. When the resonator depletes passively before the arrival of the depletion pulse, the spectrum shows a Lorentzian peak that converges to the narrow resonator spectrum as more samples are included. When the reflected pulse arrives at the detector, we see a wide-band function superimposed on the resonant peak, which converges to the spectrum of the output pulse after a time $t_w$. Clearly, the photons constituting the broader band arrived only after the pulse was detected.

The existence of this uncertainty relationship between time and frequency in the detection scheme drives home the idea that these short pulses clearly exhibit pure wave-like behaviour in these systems under the classical limit, with no information about where the photons 
are, other than within the confines of the pulse width.


\section{Conclusion and Outlook}
We have thus defined a possible physical explanation for the depletion phenomenon and provided a set of matching conditions to engineer the pulses for maximum depletion. The flexibility of the model lies in the freedom to choose a pulse shape. In prior literature, specific pulse shapes have achieved better results than others \cite{Gambetta2016}. The most important utility of this protocol is the elimination of a large bottleneck in the operation of superconducting quantum computers. There is a major reduction in the depletion time as compared to passive resonator loss. After the measurement is completed, the pulse is conditioned on the measured qubit state, taking the time $L_c/v$ to reach the resonator and $t_w$ to deplete it. Thus, when the pulses are shorter than, but comparable in their width to the length of 
the transmission line, i.e., $vt_w\lesssim L_c$, the total time it takes to deplete the resonator after measurement is $L_c/v+t_w\sim t_w$. It is practical to construct transmission lines with delay times $\sim100/\omega_r$, and with $1/\kappa\sim10^4/\omega_r$, there is potential for improving depletion times by more than a factor of $100$ (compared to current long-pulse experiments \cite{Bultink2016} with reductions by $\sim 6/\kappa$). Thus, for $\omega_r\approx10 \mathrm{GHz}$, we can reduce depletion times from $10/\kappa\approx 10\mathrm{\mu s}$ to tens of nanoseconds, now comparable to the gate times of the qubits.

The presence of higher modes in transmission line resonators has not been dealt with in this work, however a generalisation can be made to higher modes by simply coupling independently to higher frequency LC-resonators or limiting the pulse width to minimise the frequency components at higher modes. Possible extensions for this theory may include other kinds of cavities and opto-mechanical systems.

\appendix

\section{Photon number of a short Gaussian pulse}
\label{section:Appendix}
We discuss a technique to derive the photon number of each mode of a wide-band Gaussian pulse. This can easily be generalised to other pulse shapes and system-dependent eigenmode bases.

Based on the definition of photons in Multi-mode Quantum Optics, we can write the overall creation operator for the pulse as a superposition of densely-spaced discrete harmonic modes $\hat{a}^{\dagger}_{\omega_i}$:
\begin{eqnarray}
    \label{eqn:mode_coeff}
    &\hat{a}^{\dagger}_{pulse} = \displaystyle\sum_i C_i\hat{a}^{\dagger}_{\omega_i}&\\
    C_i =& \displaystyle\frac{\sqrt{\delta\omega} \left[\exp\{-(\omega_i - \omega_p)^2/4W^2\} + \exp\{-(\omega_i + \omega_p)^2/4W^2\}\right]}{\sqrt{c_n}}&
\end{eqnarray}
where $\omega_p$ is the center frequency, $W$ is the bandwidth of the pulse, $\omega_i$ is the frequency of the mode $i$, $\delta\omega$ is the difference between frequencies of successive modes and $c_n$ is the normalisation constant.

This expression comes from the Fourier transform of a pulse with a Gaussian profile. In the narrow-band regime, as is considered in the Appendix of \cite{Solgun_2013}, it is implicitly assumed that $\omega_p/W>>1$, and hence one can directly take the weight of each mode to be,
\begin{equation}
    C_i = \frac{\sqrt{\delta\omega} \exp\{-(\omega_i - \omega_p)^2/4W^2\}}{(2\pi W^2)^{1/4}}
\end{equation}
and safely integrate over just one of the Gaussians in the spectrum.

Here, we will assume $\omega_p$ to be comparable in magnitude to $W$, i.e., the pulse is short, so that there is significant contribution to the photon number from the low-frequency modes (even the 'zero-frequency' mode) due to the overlap between the two Gaussians.

The weights $C_i$ are normalised to unity in the limit of continuum of modes as,
\begin{equation}
    \begin{split}
        \sum |C_i|^2 \approx \displaystyle\frac{1}{c_n} \int_0^\infty \mathrm{d}\omega \left(e^{-(\omega - \omega_p)^2/4W^2} + e^{-(\omega + \omega_p)^2/4W^2}\right)^2 = 1\\
        \implies \displaystyle\frac{1}{c_n} \int_0^\infty \mathrm{d}\omega \left(e^{-(\omega - \omega_p)^2/2W^2} + e^{-(\omega + \omega_p)^2/2W^2} + 2e^{-\omega^2/2W^2}e^{-\omega_p^2/2W^2}\right) = 1   
    \end{split}
\end{equation}
We solve this integral by exploiting the even nature of the integrand and mirroring the limits to cover the full real line. We end up with:
\begin{equation}
    c_n = \sqrt{2\pi W^2}\left(1+e^{-\omega_p^2/2W^2}\right)
\end{equation}
The pulse is constructed with multiple modes of coherent photons, with each mode contributing an average photon number $n_i = |C_i|^2N$, where $N$ is the total number of photons in the pulse. Thus, $\displaystyle\sum_i n_i = N\displaystyle\sum_i |C_i|^2 = N$.\\
For a single mode of coherent radiation, the average energy contained in the mode is given by $n_i\hbar\omega_i$, so the total energy in the pulse is given by:
\begin{equation}
    \langle E\rangle = \displaystyle\sum_in_i\hbar\omega_i = N\displaystyle\sum_i\hbar|C_i|^2\omega_i
\end{equation}
\begin{figure}
    \centering
    \subfigure[]{
    \label{fig:PSD}
    \includegraphics[width = 0.425\textwidth]{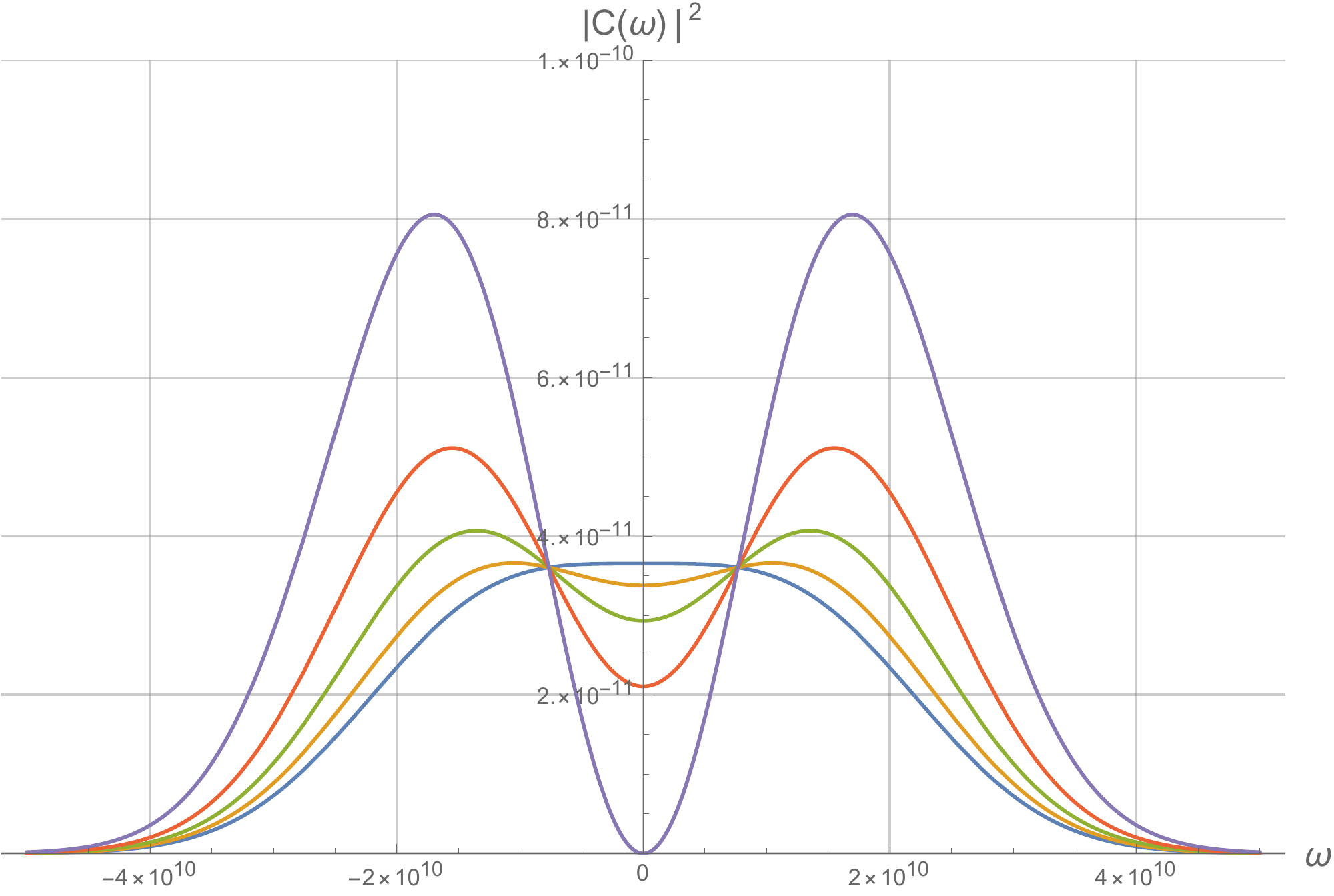}}
    \subfigure[]{
    \label{fig:Photonnum}
    \includegraphics[width = 0.48\textwidth]{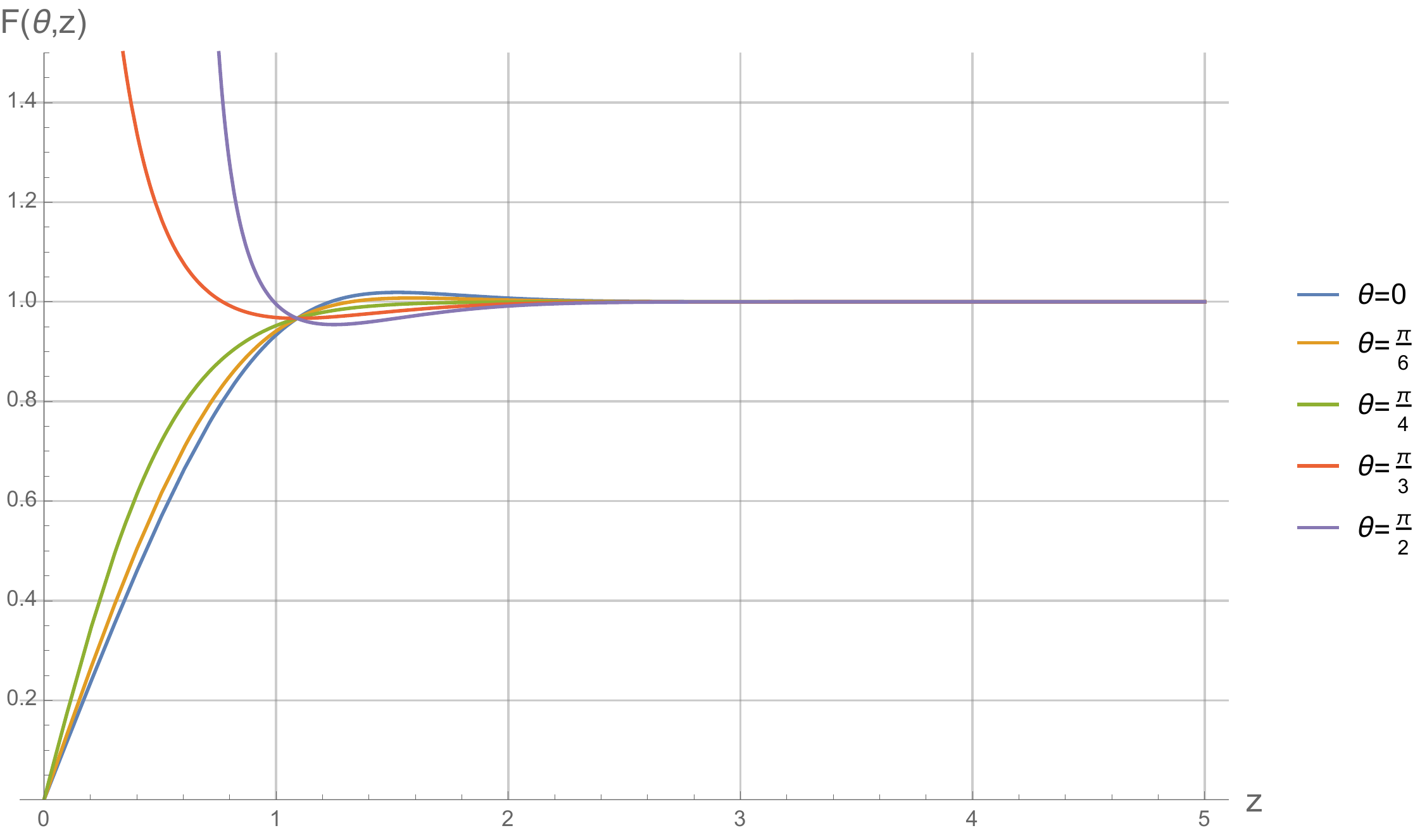}}
    \caption{a) Plot of $|C(\omega)|^2$ vs $\omega$ for a unit amplitude pulse with $\omega_p = 1.414\times 10^{10}$ and $W=1\times10^{10}$, which shows clear difference in behaviour at different CEO phases due to varying contributions from the low-frequency photons. b) Plot of $F(\theta,z)$ vs $z$ for different CEO phases. As shown, it clearly settles to the expected narrow band value for large $z$.}
    \label{fig:pulseprofile}
\end{figure}
Taking continuum of modes:
\begin{equation}
    \begin{split}
        \langle E\rangle &= \frac{N\hbar}{c_n} \int_0^\infty \omega\mathrm{d}\omega \left(e^{-(\omega - \omega_p)^2/2W^2} + e^{-(\omega + \omega_p)^2/2W^2} + 2e^{-\omega^2/2W^2}e^{-\omega_p^2/2W^2}\right)\\
        &= \frac{N\hbar}{c_n} \int_0^\infty \omega\mathrm{d}\omega \left(e^{-(\omega - \omega_p)^2/2W^2} + e^{-(\omega + \omega_p)^2/2W^2}\right) + \displaystyle\frac{2N\hbar e^{-\omega_p^2/2W^2}}{c_n} \int_0^\infty \omega e^{-\omega^2/2W^2}\mathrm{d}\omega\\
        &= \left\{\frac{2N\hbar \omega_p}{c_n}\sqrt{\frac{\pi W^2}{2}}\mathrm{erf}\left(\frac{\omega_p}{W\sqrt{2}}\right) + \frac{2N\hbar W^2}{c_n}\mathrm{exp}\left(\frac{-\omega_p^2}{2W^2}\right)\right\}+ \frac{4N\hbar W^2}{c_n}\mathrm{exp}\left(\frac{-\omega_p^2}{2W^2}\right)\\
        &= N\hbar \omega_p \left\{ \frac{\mathrm{erf}\left(\frac{\omega_p}{W\sqrt{2}}\right) + \frac{3W\sqrt{2}}{\sqrt{\pi}\omega_p}\mathrm{exp}\left(\frac{-\omega_p^2}{2W^2}\right)}{\mathrm{exp}\left(\frac{-\omega_p^2}{2W^2}\right)+1}\right\}
    \end{split}
\end{equation}
Let $z = \frac{\omega_p}{W\sqrt{2}}$,
\begin{equation}
    \implies \langle E\rangle = \displaystyle N\hbar \omega_p \left\{ \frac{\mathrm{erf}(z) + \frac{3}{z\sqrt{\pi}}\mathrm{exp}(-z^2)}{\mathrm{exp}(-z^2)+1}\right\}
\end{equation}
One can calculate the energy $\bar{E}$ of a pulse with classical methods, so it is usually a known quantity. Therefore,
\begin{equation}
    N = \frac{\bar{E}}{\hbar\omega_p} \left\{ \frac{\mathrm{exp}(-z^2)+1}{\mathrm{erf}(z)+\frac{3}{z\sqrt{\pi}}\mathrm{exp}(-z^2)}\right\}
\end{equation}
We can now generalise the expression by introducing a non-zero carrier-envelope offset (CEO) phase $\theta$ \cite{CEO}:
\begin{equation}
    \begin{split}
        N &= \frac{\bar{E}}{\hbar\omega_p}\left\{ \frac{\mathrm{cos}(2\theta)\mathrm{exp}(-z^2)+1}{\mathrm{erf}(z)+\frac{(2cos(\theta)+1)}{z\sqrt{\pi}}\mathrm{exp}(-z^2)}\right\}\\
        &= \frac{\bar{E}}{\hbar\omega_p}F(\theta, z)
    \end{split}
\end{equation}
where $\theta$ is the CEO phase. The value of $F(\theta,z)$ approaches $1$ for large $z$, as the pulse becomes longer in time and approaches the narrow-band limit. The plots for $|C_i|^2$ and $F(\theta,z)$ are given in Fig.~\ref{fig:pulseprofile}.

\nocite{*}

\bibliography{apssamp}

\end{document}